\documentclass[lettersize,journal]{IEEEtran}
\usepackage{amsmath,amsfonts}
\usepackage{algorithmic}
\usepackage{algorithm}
\usepackage{array}
\usepackage[caption=false,font=normalsize,labelfont=sf,textfont=sf]{subfig}
\usepackage{textcomp}
\usepackage{stfloats}
\usepackage{url}
\usepackage{verbatim}
\usepackage{graphicx}
\usepackage{cite}
\usepackage{color}
\usepackage{algorithm}
\usepackage{algorithmic}
\usepackage{float}
\usepackage{multirow}
\begin{document}

% \title{ChannelGPT : A Large Model for Digital Twin Channel Enabled 6G}
\title{ChannelGPT: A Large Model to Generate Digital Twin Channel for 6G Environment Intelligence}

\author{Li Yu, Lianzheng Shi, Jianhua Zhang, Jialin Wang, Zhen Zhang, Yuxiang Zhang, Guangyi Liu}

\maketitle

\begin{abstract}
6G is envisaged to provide multimodal sensing, pervasive intelligence, global coverage, global coverage, etc., which poses extreme intricacy and new challenges to the network design and optimization. As the core part of 6G, wireless channel is the carrier and enabler for the flourishing technologies and novel services, which intrinsically determines the ultimate system performance. However, how to describe and utilize the complicated and high-dynamic characteristics of wireless channel accurately and effectively still remains great challenges. To tackle this, digital twin is envisioned as a powerful technology to migrate the physical entities to virtual and computational world. In this article, we propose a large model driven digital twin channel generator (ChannelGPT) embedded with environment intelligence (EI) to enable pervasive intelligence paradigm for 6G network. EI is an iterative and interactive procedure to boost the system performance with online environment adaptivity. Firstly, ChannelGPT is capable of utilization the multimodal data from wireless channel and corresponding physical environment with the equipped sensing ability. Then, based on the fine-tuned large model, ChannelGPT can generate multi-scenario channel parameters, associated map information and wireless knowledge simultaneously, in terms of each task requirement. Furthermore, with the support of online multidimensional channel and environment information, the network entity will make accurate and immediate decisions for each 6G system layer. In practice, we also establish a ChannelGPT prototype to generate high-fidelity channel data for varied scenarios to validate the accuracy and generalization ability based on environment intelligence. 
\end{abstract}

% \begin{IEEEkeywords}
% Article submission, IEEE, IEEEtran, journal, \LaTeX, paper, template, typesetting.
% \end{IEEEkeywords}

\section{Introduction}

As increasingly innovative applications and services emerge, including multimodal sensing, autonomous driving, holographic communication, and pervasive intelligence, researchers from both industry and academia are shifting their focus to the sixth generation (6G) mobile network \cite{Transformer-Empowered}. Serving as a pivotal element of the 6G network, wireless channel is crucial determinant of performance metrics such as transmission rates, signal quality, and network capacity. To unlock the full potential of 6G mobile communication systems and accommodate the needs of these emerging services, it is essential to delve deeper into the propagation characteristics of wireless channels to achieve real-time accuracy in channel behavior assessments.

Nevertheless, with the advent of characteristic of 6G networks, such as broad frequency bands, diverse scenarios, and innovative technologies, traditional channel research methodologies have revealed significant limitations. Initially, conventional channel modeling predominantly relies on statistical analyses of channel measurements, which struggles to accurately depict channel characteristics in dynamic and heterogeneous real-world environments. Moreover, these methods exhibit a lack of generalization capability to adapt across varied scenarios. However, the characteristics of propagation channels are fundamentally influenced by physical environment factors, such as buildings, weather conditions, and the behaviors of mobile users \cite{predictive_6G_network}. Consequently, understanding the environment is crucial to accurately modeling channel fading.

Recently, digital twins have emerged as a highly promising enabler for 6G research, integrating digital environment information into communication networks to facilitate network design, simulate network performance, optimize online transmission technologies, and even interact with physical entities \cite{6G_DTN}. Infusing wireless channels with digital twin capabilities is anticipated to revolutionize 6G network design from the ground up, and is considered a cutting-edge approach. The Digital Twin Channel (DTC) is an innovative implementation in this area that accurately mirrors the characteristics of real physical channels in the digital domain \cite{wang2024DTC}. This capability allows physical communication entities to make proactive decisions based on precise virtual channel representations. DTC has the potential to deliver effective decision-making for various wireless tasks across different scenarios in 6G networks, particularly in dynamic and heterogeneous environments.

In fact, DTC leverages advanced artificial intelligence (AI) technology to capture accurate channel propagation characteristics in real time, effectively integrating digital environments with wireless communication tasks. However, many existing AI models \cite{AI_Based_zhang, E2ENet, xu2022computer_v2x, SunSemantics} are small-scale, focusing on single tasks and specific scenarios, which limits their capability to generalize across the diverse, multi-task requirements of DTC. Particularly in digital twin scenarios built on comprehensive scene maps, it is currently challenging to simultaneously make decisions on multiple tasks such as channel prediction, resource allocation, and interference management, etc.

At present, with the advancement of hardware and the surge in data size, large models have demonstrated excellent generalization capabilities and efficient processing capabilities for complex tasks in the fields of natural language processing and computer vision. The introduction of the Transformer architecture and its self-attention mechanism significantly enhances the capability to capture long-range dependencies, while making the model training process more efficient through parallelization\cite{vaswani2023attentionneed}. This technological advancement has promoted the development of large models, especially OpenAI's GPT series models, which have demonstrated their potential in multimodal tasks.

With the remarkable success of large models, their application in the field of communications has gradually attracted attention. Although the application of large models in the field of communications is still in its early stage, recent preliminary explorations on this issue\cite{Big_Zhaoyang2024, LargeModel_Telecom2024, LLM4CP} show great potential in this direction. However, the problem of how to effectively combine and run large models under the DTC framework has not been solved. Therefore, we propose a channel generation pre-trained transformer (ChannelGPT) to support the DTC framework. Taking advantage of the DTC framework, ChannelGPT can more accurately characterize complex channel fading information and provide efficient and intelligent decision support, improving the overall performance and responsiveness of 6G networks. 

We believe that integrating large model technology into DTC will help realize environment intelligence (EI), which includes not only simple perception of the physical environment, but also a comprehensive understanding and dynamic response to environment changes, user needs, channel characteristics, etc., making the decision process more complex and improving the adaptability and intelligence level of 6G networks.

\section{DTC enabled 6G by environment intelligence}
\begin{table*}[htbp]
\setlength{\tabcolsep}{11pt} %%%
\renewcommand\arraystretch{1.2}  %% 
\caption{Comparative analysis of DTC based on different methods about modeling capability, cost, generalization, data source, and multi-task processing.}
\label{related_work_summary_table}
\begin{center}
\begin{tabular}{|m{1.5cm}<{\centering}|m{0.7cm}<{\centering}|m{3.4cm}<{\centering}|m{1.8cm}<{\centering}|m{2cm}<{\centering}|m{1.8cm}<{\centering}|m{1.3cm}<{\centering}|}
\hline
 \textbf{Category} & \textbf{Ref.} & \textbf{Modeling capability} & \textbf{Cost} & \textbf{Generalization} & \textbf{Data source} & \textbf{Multi-task processing} \\
\hline
 Traditional method & \cite{3GPP_TR38901} & Mathematical representation & Model formula calculation & Specific scenario & Channel measurements & ×  \\ 
\hline
 \multirow{3}{*}[-1.2em]{Small models} & 
 \cite{E2ENet} & 
 Immune extra error & 
 MT & 
 Specific scenario & Simulation dataset & ×  \\ 
 \cline{2-7}
 & \cite{xu2022computer_v2x} & Avoid pilot overhead  & MT & 
 Specific scenario & 
 Simulation dataset & 
 ×  \\ 
 \cline{2-7}
 & \cite{SunSemantics} & Highly efficient  beam prediction & MT & 
 Specific scenario & 
 Simulation dataset & × \\
\hline
\multirow{3}{*}[-1.7em]{Large models} 
 & \cite{Big_Zhaoyang2024} & 
 High-efficient, sustainable, versatile, and extensible wireless intelligence & 
 HD \& MT & 
 Diverse scenarios & 
 Simulation \& measurement & 
 \checkmark \\
 \cline{2-7}
 & \cite{LargeModel_Telecom2024} & 
 Self-evolving network & 
 HD \& MT & 
 Diverse scenarios & 
 Simulation \& measurement & 
 \checkmark \\
 \cline{2-7}
 & \cite{LLM4CP} & 
 High accuracy and generalization capability & 
 HD \& MT & 
 Diverse scenarios & 
 Simulation \& measurement & 
 \checkmark \\
 \cline{2-7}
\hline
\multicolumn{7}{|l|}{HD: Hardware devices; MT: Model training.}\\
\hline
\end{tabular}
\end{center}
\end{table*}
%%%%%%%%%%%%%%%%%%%%%%%%%%%%%%%%%%%%%%%%%%%%%%%%%%%%%%%%%%%%%%%
% 在本节中，我们介绍了DTC的工作流程和环境智能，并详细阐述了DTC与6G网络、DTC与ChannelGPT的完整交互过程。其次，将ChannelGPT集成到DTC框架中赋能DTC。
In this section, we introduce the DTC workflow with EI. Additionally, the complete interaction process between DTC and the 6G network, as well as between DTC and ChannelGPT, is explained in detail. Besides, ChannelGPT is integrated into the DTC framework. 

\begin{figure*}[htbp]
\centerline{\includegraphics[scale=0.3]{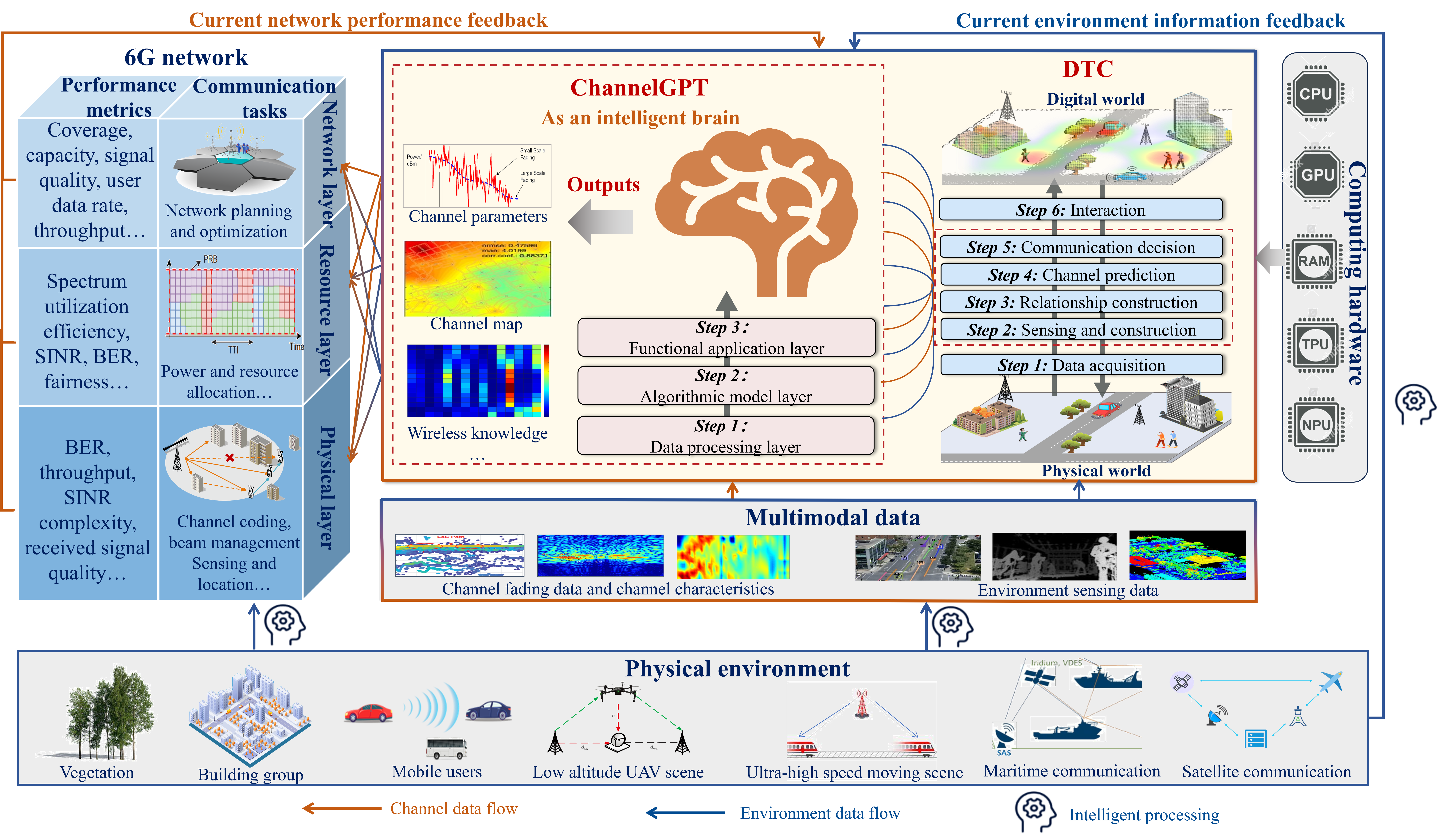}}
\caption{ChannelGPT enabled 6G architecture with EI.}
\label{DTC_6G}
\end{figure*}

\subsection{DTC and 6G Network}
\subsubsection{DTC workflow}
% 这一小节简单介绍一下DTC的六个步骤，简单说一下每一步的功能和目的即可
% 如图所示，DTC实现了无线信道和通信决策从物理世界到数字世界的在线映射，是使能6G网络有潜力的方法。
As shown in Fig. \ref{DTC_6G}, DTC enables the online mapping of wireless channel and communication decisions from the physical world to the digital world, representing a promising approach for the 6G network. 
% DTC的工作流程分为六个步骤。首先，step1旨在从物理世界采集多模态数据，其中包括了信道数据和环境感知数据，确保DTC能够准确反映实际环境状况。
The workflow of DTC is divided into six steps. Step 1 aims to acquire multimodal data from the physical environment, including channel data and environment sensing data, ensuring that DTC accurately reflects the actual environment conditions. 
% step 2分析采集到的多模态数据以重建数字世界。step 3中对信道和环境感知数据进行处理，分析环境与信道之间的潜在关系。
Step 2 involves analyzing the collected multimodal data to reconstruct the digital world. In Step 3, the channel data and environment sensing data are processed to analyze the relationships between the environment and the channel. 
% step 4根据构建的关系，通过AI来挖掘环境与信道之间的深层映射，预测信道衰落信息。
In Step 4, the established relationships are utilized to leverage AI for digging out deep mappings between the environment and the channel, predicting channel fading information. 
% step 5基于预测结果的通信策略，这是DTC赋能6g网络的关键一步，满足6G网络各层级的决策需求。
In Step 5, communication decisions are developed based on the prediction results. This is a key step in enabling the 6G network, meeting the decision-making needs across various layers of the network. 
% step 6实现DTC与6G网络的在线交互，通过在线反馈机制优化模型和决策。
In Step 6, online interaction between DTC and 6G network is achieved through a feedback mechanism that optimizes models and decisions. 
\subsubsection{Environment intelligence}
% 环境智能不仅作为DTC和6G网络的基础能力，也是6G网络与DTC交互过程中的核心元素。
EI serves not only as a foundational capability for both DTC and 6G network but also as a central element in the interaction process between them. 
% 环境信息流的递进和交互过程如图所示，从物理世界采集多模态数据以支撑DTC和6G网络的数据需求。
% 环境信息流的递进和交互过程如图所示
% The progressive interaction of environment information flow is illustrated in Fig. \ref{DTC_6G}, demonstrating the collection of multimodal data from the physical environment to support the data requirements of both DTC and 6G network. 
The progressive interaction of environment information flow is illustrated in Fig. \ref{DTC_6G}. 
% The progressive and interactive process of environment information flow is enabled by the large model, as shown in Fig. \ref{DTC_6G}. 
% 环境智能首先体现在从物理世界采集的无线环境信道进行智能处理，如通过信道特性处理算法得到大小尺度参数、基于环境数据的AI感知算法获得图像或者点云信息中感兴趣的目标信息等。进一步，还可以通过物理规律构建无线环境知识库，得到更深刻的环境认知。环境智能后续挖掘环境与信道的映射关系提供了更深入的先验知识。
EI is first manifested through the intelligent processing of wireless environment channel collected from the physical environment. This includes deriving large-scale and small-scale parameters using channel characteristic processing algorithms and obtaining target information of interest from images or point cloud data through AI perception algorithms based on environment data. Furthermore, the radio environment knowledge pool (REKP) can be constructed using physical principles, leading to a deeper understanding of the environment \cite{wang2023REKP}. EI provides valuable prior insights for exploring the mapping relationship between the environment and the channel. 
% 其次，环境智能实现了DTC和6G网络的在线交互。具体来说，DTC具有提供无线环境图和知识的能力，为6G网络决策提供有效参考。其中，无线环境图和知识代表了无线信道参数、信道图和无线知识。同时，根据6G网络性能指标质量，物理环境会根据需求及时采集反馈给DTC。这一反馈机制使得DTC能够重新调整决策，优化网络配置，从而提高整体系统的智能化水平。
% The wireless channel and environment knowledge contains channel parameters, channel map, and wireless knowledge.
Moreover, EI achieves the online interaction between DTC and 6G network. Specifically, DTC has the capability to provide channel parameters, channel map, and wireless knowledge, offering effective references for decision-making within 6G network. Simultaneously, based on the quality of performance metrics from the 6G network, the physical environment can timely collect and feedback data to DTC as needed. This feedback mechanism enables DTC to readjust decisions and optimize network configurations, thereby enhancing the overall level of system intelligence. 

\subsubsection{Online interaction between DTC and 6G network}
% 这一小节主要是围绕6G网络和DTC之间的交互流程，如何赋能，如何反馈，如何更新
% 如图所示，通过闭环的在线反馈机制，DTC能够为6G网络提供在线支持，而6G网络通过性能指标在线评估反馈给DTC优化模型和决策。这种在线反馈机制确保了6G网络在高动态的、多样的场景中的高效运作。
As shown in Fig. \ref{DTC_6G}, through a closed-loop online feedback mechanism, DTC can provide online support for the 6G network, while the 6G network assesses performance metrics and feeds back to DTC to optimize models and decisions. This online feedback mechanism ensures the efficient operation of the 6G network in high-dynamic and diverse scenarios. 
% 具体来说，将6G网络自底向上划分为三个层级，分别为物理层，资源层，网络功能层，DTC根据特定需求来使能对应层。
Specifically, the 6G network is hierarchically divided into three layers from the bottom up: the physical layer, resource layer, and network function layer. DTC enables each layer based on its specific requirements. 
\subsubsection*{\textbf{Physical layer}}
% 在6G 网络中，物理层常见技术包括有：预编码技术，信道编码，调制，波束管理等等。这些技术的有效实施依赖于有效的无线环境图和知识。例如，DTC可以为预编码技术提供在线的信道状态信息、路径损耗、位置信息等，以优化发送信号的配置，确保在复杂环境中实现高效的信号传输。对于波束管理，DTC能够分析用户的位置信息和移动轨迹，从而动态调整波束码本和方向，优化信号覆盖和增强用户体验。
In the 6G network, common technologies at the physical layer include precoding techniques, channel coding, modulation, and beam management, among others. The implementation of these technologies relies on channel parameters, channel map, and wireless knowledge. For example, DTC can provide online channel state information (CSI), path loss, location information and other data which can be considered a form of channel parameters and channel map for precoding techniques to optimize the configuration of transmitted signals, ensuring efficient signal transmission in complex environments. For beam management, DTC can analyze user location information and movement trajectories to dynamically adjust the beam codebook and direction which can be considered a form of wireless knowledge, optimizing signal coverage and enhancing user experience. 
% 物理层需要通过一些关键指标来评估，比如：信干噪比、接收信号质量、误码率和波束赋形增益等，并将这些评估结果反馈给DTC实现交互。
The physical layer needs to evaluate key metrics such as signal-to-interference-plus-noise ratio (SINR), received signal quality, bit error rate (BER), and beamforming gain, and feedback these assessment results to DTC for interaction. 
\subsubsection*{\textbf{Resource layer}}
% 资源层负责高效管理和分配网络资源，以满足不同用户的需求。
The resource layer is responsible for efficiently managing and allocating network resources to meet the diverse needs of users. 
% 例如，在功率分配方面，DTC能够提供路损、用户位置等信息，资源层动态调整每个用户的发射功率，以降低干扰并优化信号质量。在频谱资源分配方面，DTC可以提供CSI、SNR等信息，资源层动态分配频谱资源，确保资源的高效利用和公平性。
For example, in power allocation, DTC can provide information such as path loss and user location which can be considered a form of channel map, enabling the resource layer to dynamically adjust the transmission power for each user, reducing interference and optimizing signal quality. In spectrum resource allocation, DTC can supply data like CSI and SNR, allowing the resource layer to dynamically allocate spectrum resources, ensuring efficient utilization and fairness. 
% 资源层通过如频谱利用率、延迟、总功耗等指标来评估性能。这些评估结果将反馈给DTC，以实现交互并优化资源分配策略。
The resource layer evaluates its performance using metrics such as spectrum utilization, latency, and total power consumption. These assessment results will be fed back to DTC to enable interaction and optimize resource allocation strategies. 
\subsubsection*{\textbf{Network function layer}}
% 网络功能层负责实现多种网络服务和功能，包括网络规划、网络优化等。
The network function layer is responsible for implementing various network services and functions, including network planning and optimization. 
% 网络功能层更倾向于使用channel map和无线知识来满足任务需求。
The network function layer tends to use channel map and wireless knowledge to meet the communication task requirements.
% 例如，在网络规划阶段，DTC提供路径损耗、SNR等。通过分析这些数据，网络功能层能够评估不同配置方案的可行性，从而制定出合理的基站覆盖方案。这一过程不仅确保了网络在覆盖范围内的有效性，还在资源使用和成本方面寻求最佳平衡，确保网络的经济性和可持续发展。
% 而在网络优化过程中，DTC提供RSRP、power、基站信息等。网络功能层可以动态调整基站的发射功率，以优化信号覆盖和减少干扰，提升用户的连接质量和满意度。

For example, in the network planning phase, DTC provides path loss, SNR, base station (BS) information, and other data. By analyzing these information, the network function layer can evaluate the feasibility of different configuration schemes and develop reasonable BS coverage plans. This process ensures network effectiveness within coverage areas while seeking an optimal balance in resource utilization and costs, thereby ensuring the network's economy and sustainability. In the network optimization process, DTC provides reference signal receiving power (RSRP), power, and other metrics, enabling the network function layer to dynamically adjust the transmission power of BS to optimize signal coverage and reduce interference, thereby enhancing user connection quality and satisfaction. 
% 网络功能层通过如覆盖范围、容量、用户速率等指标来评估性能。这些评估结果将反馈给DTC，保持网络的稳定性和高效性。
The network function layer evaluates its performance using metrics such as coverage area, capacity, and user data rates. These assessment results will be fed back to DTC to maintain the network's stability and efficiency. 

\subsection{DTC and ChannelGPT}
% 这一小节，简单讲一下两者之间的关系，以及各自的定位，负责什么作用，以及ChannelGPT所需的算力支撑和数据支撑等。

% 在6G网络中，DTC框架亟需一个具备强大的智能核心，处理并融合多模态数据，执行通信任务，因此我们提出channel GPT。
In 6G network, the DTC framework is in dire need of an intelligent core to process and fuse multimodal data, and perform communication tasks. ChannelGPT is a promising implementation. 

% ChannelGPT是DTC智能大脑，负责从DTC提供的数据中进行复杂的建模、生成、预测和优化。DTC是ChannelGPT的数字驱动平台，提供了用于建模和学习的数据，以及反馈和验证的平台。ChannelGPT赋能DTC工作流程步骤2到5，为整个流程提供智能支持。
ChannelGPT is the DTC Intelligent Brain, responsible for complex modeling, generation, prediction, and optimization from the data provided by DTC. DTC is the digitally driven platform for ChannelGPT, providing the data for modeling and learning and the platform for feedback and validation. ChannelGPT empowers the DTC workflow from Steps 2 to 5, providing intelligent support for the entire process. 
% 此外，ChannelGPT赋能DTC也需要强大的硬件资源，CPU,GPU,RAM,TPU,NPU等，稍微展开说一下。
% 此外，ChannelGPT的高效运作需要强大的硬件资源，包括CPU、GPU、RAM、TPU和NPU等，以确保其处理海量的多模态数据并进行复杂的模型训练和推理。
Additionally, the efficient operation of ChannelGPT requires robust hardware resources, including CPU, GPU, RAM, TPU, and NPU. These resources are essential for handling large volumes of multimodal data and conducting complex model training and inference.

\section{ChannelGPT as a key enabler for DTC}
% 本节详细阐述了在DTC下ChannelGPT的架构和能力，其架构分层如图所示。
This section details the overall architecture and abilities of ChannelGPT under DTC. Its architectural layering is shown in Fig. \ref{framework_gpt}. 

\subsection{Framework}
\begin{figure*}[t]
\centerline{\includegraphics[scale=0.45]{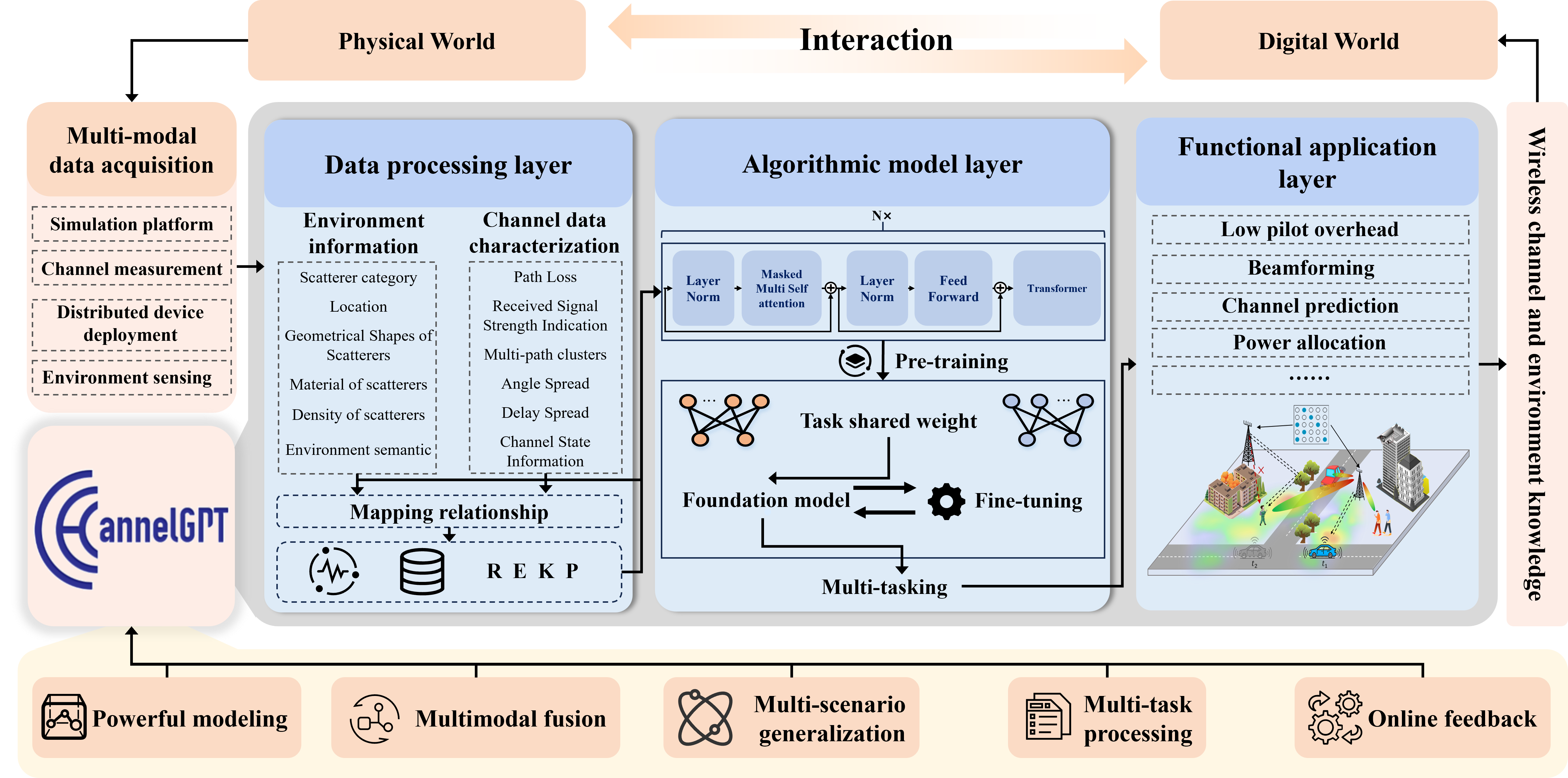}}
\caption{The framework of ChannelGPT.}
\label{framework_gpt}
\end{figure*}

% 为了服务于DTC运作，ChannelGPT被设计为三层核心架构，包含数据层、算法建模层和功能应用层。
To support the operation of DTC, ChannelGPT is designed with a three-layer core architecture, containing the data processing layer, algorithm modeling layer, and functional application layer, as shown in Fig. \ref{framework_gpt}.

\subsubsection{Data processing layer}
% 数据层是ChannelGPT框架的基础层，负责收集和处理来自DTC平台的海量的多模态数据，其中主要数据类型包括环境信息和信道数据。在海量数据集初步构建期间，通过ray-tracing仿真或实际信道测量获取信道数据。其次，可利用carla、AirSim等自动驾驶仿真平台或不同传感器获取环境数据。在时间、空间不同维度上对齐信道与环境数据。此外，具体实施阶段可通过分布式设备来更新、补充数据集，这需要考虑到历史数据管理、分布式实时数据的调度、缓存和同步等问题。
The data processing layer serves as the foundational component of the ChannelGPT framework, collecting and processing vast amounts of multimodal data from the DTC platform. The primary data types include environment information and channel data. During the initial phase of large-scale dataset construction, channel data can be obtained through ray-tracing simulation or channel measurement. Additionally, autonomous driving simulation platforms like CARLA, AirSim, or various sensors can be employed to acquire environment data. The dataset is aligned in different temporal and spatial dimensions to ensure coherence between channel and environment information. In the implementation phase, distributed devices can collect data and update dataset, which requires consideration of historical data management, distributed online data scheduling, caching, and synchronization. 
% 正如[]中所评论的，仅依靠原始数据作为模型训练的数据集效果可能是差强人意的，因为原始数据存在着冗余和噪声，因此，REKP是ChannelGPT的有力推动者，挖掘无线电传播与其环境之间的本质、关系和规律模式是有重要意义的。
As commented in \cite{wang2023REKP}, relying only on raw data as a dataset for model training may be poorly effective because of redundancy and noise in the raw data. Therefore, REKP is a strong promoter of ChannelGPT. It is of great significance to dig out the nature, relationships, and patterns between radio propagation and the environment. 

\subsubsection{Algorithm modeling layer}
% 算法建模层是ChannelGPT的核心，负责基于数据层提供的多模态数据进行建模和学习。
The algorithm modeling layer is the core of ChannelGPT, responsible for modeling and learning based on the multimodal data provided by the data processing layer.
% 这一层中需要关注的几点：1、现有的大模型算法是地基；2、突出结合信道知识（REKP）设计信道专属的大模型是具有重要意义的；3、预训练和微调；4、算法设计要关注DTC的需求，多任务共享权重机制

% 现有的大模型架构在处理复杂的多模态数据方面具备良好的表现，如GPT、LLaMA（Large Language Model Meta AI）、 Mix of Experts (MoE)等。通过借鉴这些现有大模型架构，ChannelGPT得以快速形成初步的框架，为处理海量的信道数据和环境数据打下坚实基础。
Firstly, existing large model architectures have good performance in handling complex multimodal data, such as GPT, Large Language Model Meta AI (LLaMA), Mix of Experts (MoE) and so on \cite{yang2024harnessing}. With these existing large model architectures, ChannelGPT can quickly form a preliminary framework, laying a solid foundation for processing massive channel data and environment data. 
% 然而，ChannelGPT的优势之一在于其结合了REKP来设计专门针对信道的大模型。通过将信道传播的物理原理融入到模型架构中，ChannelGPT不仅可以更精确地预测信道衰落状态，还可以有效提高DTC对信道环境的适应性。
% However, one of the advantages of ChannelGPT lies in its integration of the REKP to design a large model specifically for wireless channel. By incorporating the physical principles of channel propagation into the model architecture, ChannelGPT not only enhances the precision of channel fading prediction but also significantly improves the adaptability of DTC to varying channel environments. 
One of the advantages of ChannelGPT lies in its integration of the REKP to design a large model specifically for wireless channel. By incorporating the physical principles of channel propagation into the model architecture, ChannelGPT not only enhances the precision of channel fading prediction, but also significantly improves the adaptability of DTC to varying channel environments. 

% 其次，预训练和微调是ChannelGPT核心的学习机制。DTC框架可以提供海量的信道数据、环境数据，这些数据对于模型理解信道传播机制至关重要。ChannelGPT在大规模的多模态数据集上进行预训练。在此基础上，ChannelGPT也可通过微调机制进一步调整模型参数，使其实现最佳表现为通信任务需求。
Secondly, pre-training and fine-tuning are the core learning mechanisms of ChannelGPT. the DTC can provide massive channel data and environment data, which are crucial for the model to understand the channel propagation mechanism. ChannelGPT performs pre-training on large-scale multimodal datasets. Based on the pre-trained model, ChannelGPT can further optimize the model parameters through the fine-tuning mechanism to achieve optimal performance according to communication tasks demand.

% 此外，6G引入了许多新技术和应用场景，例如超大规模天线阵列、智能表面（如智能反射表面，IRS）、感知通信一体化等，ChannelGPT 的多任务处理能力显得尤为重要。ChannelGPT需要在同一框架内处理多种信道相关任务，如信道预测、波束成形、干扰管理、资源分配等。
Thirdly, 6G introduces increasing novel technologies and application scenarios, such as extremely large-scale multiple-input-multiple-output (XL-MIMO), reconfigurable intelligent surface (RIS), and integrated sensing and communication (ISAC). 
% Therefore, the multitasking capability of ChannelGPT is particularly crucial. 
ChannelGPT handles a variety of wireless channel tasks such as channel prediction, beamforming, resource allocation, and more. 
% 由于信道预测任务与波束成形任务之间可能存在相关性，准确的信道预测可以帮助优化波束成形策略。因此，ChannelGPT通过任务关系建模，能够捕捉不同任务之间的潜在相关性，动态调整信息共享比例，实现任务间的协同优化，从而提高整体系统的性能。
There is a certain correlation between different wireless channel tasks, for example, accurate channel prediction can help optimize beamforming strategies. 
% There may be correlations between channel prediction and beamforming tasks, where accurate channel prediction can help optimize beamforming strategies. 
Therefore, through task relationship modeling, ChannelGPT can capture potential correlations between different tasks and dynamically adjust the proportion of information sharing. This enables collaborative optimization between tasks, thereby enhancing the overall system performance. 
% ChannelGPT 使用共享表示学习来最大化通信任务之间的知识共享。不同任务可能共享底层的特征表示，也可能需要各自特定的高层特征。通过设计自适应的任务共享机制，ChannelGPT 能够在共享基础上保持任务特异性的表示，从而在处理不同任务时实现更高的效率和精度。
% ChannelGPT 使用共享表示学习来最大化通信任务之间的知识共享。不同任务可能共享底层的特征表示，也可能需要各自特定的高层特征。ChannelGPT 能够在共享基础上保持任务特异性的表示，从而在处理不同任务时实现更高的效率和精度。
ChannelGPT utilizes shared representation learning to maximize knowledge transfer between communication tasks. Different tasks share low-level feature representations and require task-specific high-level features. Then it maintains task-specific representations on a shared foundation, thereby achieving higher efficiency and accuracy when handling various tasks. 
\subsubsection{Functional application layer}
% 功能应用层是ChannelGPT的顶层结构，将得到的无线信道和环境知识作用于6G网络，以满足网络不同层次的功能需求。
% The function application layer is the top layer of ChannelGPT. It translates the theoretical advancements of the algorithm modeling layer into practical application functionalities, ensuring that advanced modeling and learning capabilities are effectively implemented in DTC systems. 
% 功能应用层是ChannelGPT的顶层，将获取的无线信道和环境知识应用到6G网络中，满足不同网络层的功能需求。无线信道和环境知识包括信道参数、信道映射和无线知识。
The functional application layer is the top layer of ChannelGPT, where the acquired wireless channel and environment knowledge are applied to the 6G network to meet the functional requirements of different network layers. The wireless channel and environment knowledge contains channel parameters, channel map, and wireless knowledge. 

% 随着6G新技术不断涌现，DTC系统的功能需求会越来越多。面向任务的接口或插件设计允许系统根据不同的应用场景和需求进行灵活配置和扩展，更便于满足用户个性化需求。
As new technologies continue to emerge in 6G, DTC functional requirements will be more and more. Task-oriented interface or plugin designs allow the system to be flexibly configured and extended according to different application scenarios and requirements. This flexibility makes it easier to meet the personalized needs of users.
% 其次，信道衰落状态受各种因素影响，如用户移动性、环境变化等，呈现出动态变化性，这决定了模型必须具备自适应调整的能力。因此，单靠静态模型难以应对这些不断变化的条件。ChannelGPT 利用分布式节点的信息采集实时的多模态数据，将这些不同模态的数据进行融合，从而全面理解和处理复杂的信道衰落信息。ChannelGPT 通过动态自适应调整机制，根据实时数据中的变化自动优化模型，确保其在新的环境条件下保持高效和准确。
Additionally, the channel fading state is influenced by various factors, such as user mobility and environment changes, exhibiting dynamic variability. This makes it crucial for the model to possess adaptive adjustment capability. Therefore, relying solely on static models is insufficient to cope with these constantly changing conditions effectively. ChannelGPT leverages distributed nodes to collect online multimodal data and integrate these diverse data modalities to comprehensively understand and process complex channel fading information. Based on it, ChannelGPT employs a dynamic adaptive adjustment mechanism that automatically optimizes the model based on changes in online data, ensuring high efficiency and accuracy under new environment conditions. 

\subsection{The capabilities of ChannelGPT}
% 作为DTC的智能核心，ChannelGPT具备了一系列关键能力，这些能力使其能够应对6G通信系统的复杂挑战。我们将深入探讨ChannelGPT框架所展现出的五大核心能力：强大的建模能力、多模态融合、多场景泛化能力、多任务处理能力以及实时反馈机制。
As the intelligent core of the DTC framework, ChannelGPT possesses the key capabilities that empower it to tackle the challenges of the 6G communication system in the future. The following sections delve into the five core capabilities of the ChannelGPT framework: powerful modeling capability, multimodal fusion, Multi-scenario generalization, multi-task processing capability, and online feedback mechanism. 
% 以下是 ChannelGPT 的核心能力：
% The following are the core capabilities of ChannelGPT:
\subsubsection{Powerful modeling capability}
% ChannelGPT 通过大规模参数和先进的模型架构，能够在大量感知数据、信道数据中学习复杂的映射关系。参数量规模优势使得 ChannelGPT 能够捕捉到细粒度的数据特征，实现高精度的信道表征与预测。例如，在进行信道状态信息(CSI)序列预测任务时，ChannelGPT能够以更先进的性能实现根据历史CSI预测未来时刻的CSI，以减少大规模多输入多输出（m-MIMO）系统中反馈或估计开销。
With large-scale parameters and advanced architecture, ChannelGPT can learn complex mappings from vast amounts of sensing and channel data. The advantage of parameter scale enables it to capture fine-grained data features, achieving highly accurate channel representation and prediction. For example, for CSI sequence prediction, ChannelGPT delivers superior performance by predicting future CSI based on historical data, effectively reducing feedback or estimation overhead in a massive MIMO system.

\subsubsection{Multimodal fusion}
% 不同的物理条件会对无线信号传播产生不同的影响。单一模态的数据（例如仅依赖信道数据）难以全面反映信道变化的动态特性。因此，为了准确捕捉信道特性，适应动态的信道环境，多模态数据融合能力变得尤为关键。
Varying physical conditions have different impacts on wireless signal propagation. Single-modal data is insufficient to fully capture the dynamic nature of channel variations, such as only relying on channel data. Therefore, to accurately characterize channel properties and adapt to dynamic environments, multimodal data fusion capability is essential. 
% ChannelGPT能够有效融合多模态数据，比如信道数据、环境感知数据等，通过跨模态对齐和模态间信息增强机制，可以最大化不同数据源之间的协同作用。例如，多视角图像和CSI的联合使用可使ChannelGPT更好地理解数据的映射关系，从而提升信道预测的精度。
ChannelGPT can effectively integrate multimodal data, such as channel data and environment data. Through cross-modal alignment and inter-modal information enhancement mechanism, it maximizes the synergy between different data sources. For example, the combined use of multi-view images and CSI enables ChannelGPT to understand data mapping relationships deeply, improving the accuracy of channel prediction. 

\subsubsection{Multi-scenario generalization}
% 场景多样性对6G通信系统的稳定性、可靠性提出了巨大挑战。ChannelGPT 可以在不同的场景中泛化处理，例如3GPP标准中的城市宏观环境（UMa）、城市微观环境（UMi）、远程大尺度环境（RMa）以及室内环境等。ChannelGPT 可以根据新场景调整其模型参数，保持模型性能的稳定性和准确性。（这块不太会描述）
The diversity of scenarios poses significant challenges to the stability and reliability of 6G communication systems. ChannelGPT can generalize in various scenarios defined by 3GPP standards, such as the macrocell (UMa), urban microcell (UMi), rural macro (RMa) \cite{3GPP_TR38901}, and other self-defined scenarios. ChannelGPT can adjust its model parameters to new scenarios, maintaining the stability and accuracy of model performance. 

\subsubsection{Multi-task processing}
% ChannelGPT的多任务学习能力使其能够在单一框架内高效处理多种信道任务。通过多任务学习，模型可以在任务间实现信息的高效共享与协同优化。
Multi-task processing capability enables it to efficiently handle various communication tasks within a single framework. 
% 通过利用多任务学习，ChannelGPT促进了不同任务之间有效的信息共享和协作优化。这种方法不仅减少了计算冗余，而且通过利用共享的底层特征提高了整体性能，允许模型在6G通信系统中同时跨多个任务进行动态调整和优化。
By leveraging multi-task learning, ChannelGPT facilitates effective information sharing and collaborative optimization between different tasks. This approach not only reduces computational redundancy but also enhances the overall performance by utilizing shared low-level features, allowing the model to adjust and optimize multiple tasks simultaneously in 6G communication systems. 
\subsubsection{Online feedback mechanism}
% ChannelGPT具备实时反馈能力，这一机制显著增强了其在动态信道环境中的适应性和智能化水平。在DTC中，当模型无法根据最新的多模态数据及时做出调整，将导致预测失误和性能下降。因此，实时反馈机制成为确保ChannelGPT高效运行的关键。
ChannelGPT is equipped with online feedback capability, which significantly enhances its adaptability and intelligence in dynamic environments. 
% In the DTC system, where channel states and external conditions change rapidly, failure to adjust the model based on the latest sensing information can lead to inaccurate predictions and performance degradation. 
In DTC, when the model is unable to make timely adjustments based on the latest multimodal data, it will lead to prediction errors and performance degradation. 
Therefore, the online feedback mechanism is essential for ensuring that ChannelGPT operates efficiently, maintaining high performance in a dynamic environment. 
% ChannelGPT能够及时获得最新的多模态数据信息，使其能够依据新数据对内部参数进行自适应调整，确保信道预测、波束成形等任务的准确性和时效性。
ChannelGPT can promptly obtain the latest multimodal data. This feedback allows ChannelGPT to adaptively adjust its internal parameters, ensuring the accuracy and timeliness of tasks. 

\section{Case Studies}
% 在本节中，我们利用GPT2在CSI时间序列预测和基于多模态信息的信道预测任务上进行了仿真验证，以证明ChannelGPT的优势和可行性。
In this section, we demonstrate the advantages and feasibility of ChannelGPT developed from GPT2 for the verification on CSI time series prediction and multimodal information-based multi-scenario channel prediction.  
\subsection{CSI prediction}
\subsubsection{Dataset construction}
% 我们使用图*的场景来构建数据集来支撑仿真验证。考虑SISO系统中采用OFDM调制，其中子载波个数为69，Tx放置在场景中间的建筑物顶部，一系列 Rx 放置在街道中间进行 RT 模拟，设置好移动端运动轨迹，构建CSI时间序列数据集。
We construct the dataset for simulation validation as shown in Fig.\ref{scenarios_performance_curve} (a). Consider orthogonal frequency division multiplexing (OFDM) modulation in the single-input single-output system, with 69 subcarriers. The transmitter (Tx) is placed on the rooftop of a building in the center of the scenario, while a series of receivers (Rx) are placed in the middle of the street for ray-tracing (RT) simulations. The mobile user's trajectory is predefined to generate the CSI time-series dataset. 
\subsubsection{Model training and performance evaluation}
\begin{figure}[htbp]
\centerline{\includegraphics[scale=0.65]{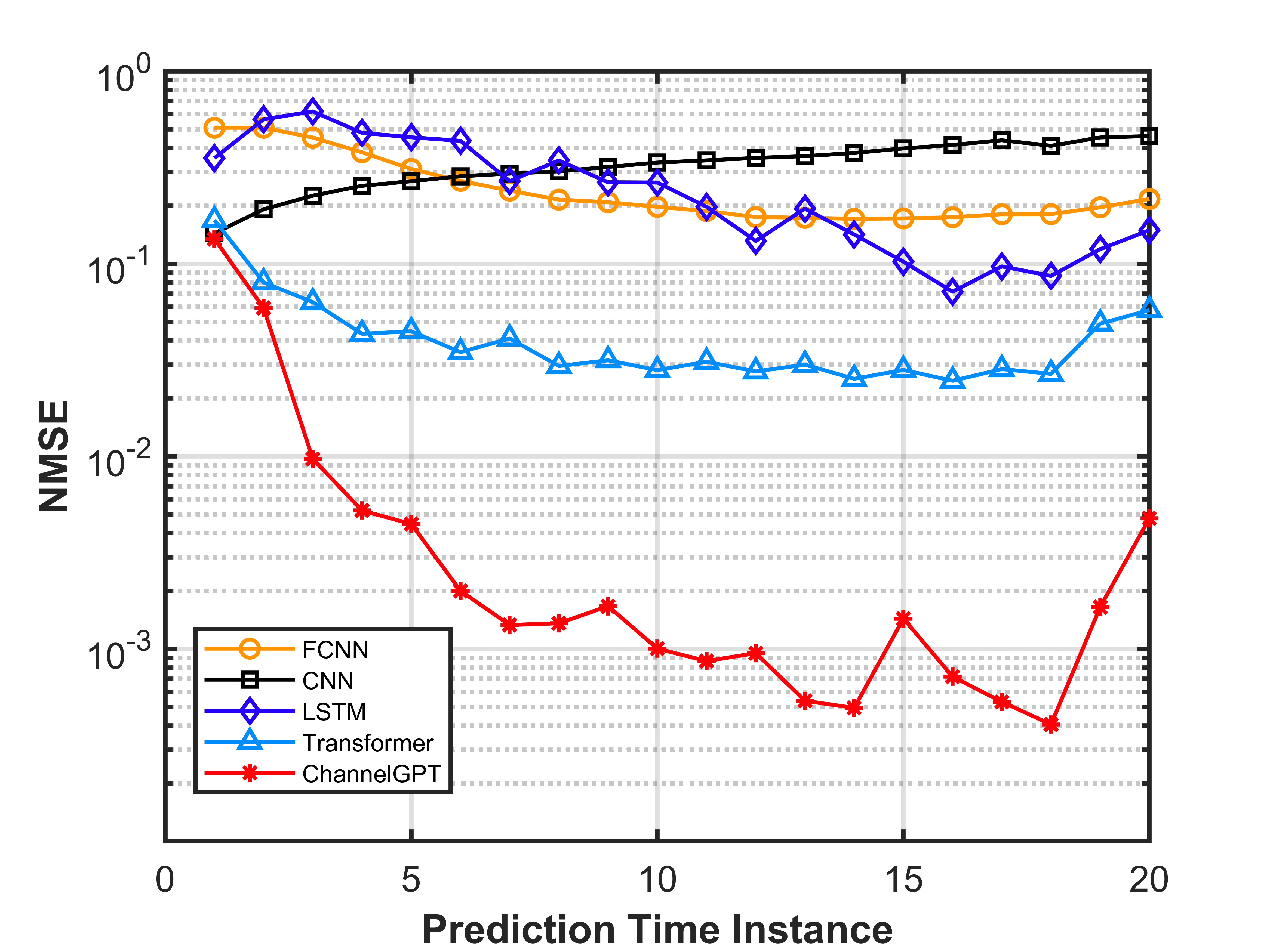}}
\caption{\textcolor{black}{The NMSE performance of ChannelGPT and other baselines versus different
time instance for long-term channel prediction.}}
\label{test_1_20_loss}
\end{figure}

% 使用构建的CSI时间序列数据集来训练模型，以减少导频资源开销。
The constructed CSI time series dataset is used to train the model to reduce the guide frequency resource overhead. 
% 具体来说，通过历史25个时隙的CSI预测未来N个连续时隙的CSI，其预测时隙范围为1-25。我们选择FCNN模型作为基线。为了评估性能，我们将真实CSI和预测CSI的归一化均方误差 (NMSE)作为测试集loss，分别对比FCNN与ChannelGPT在该预测任务下的步长与对应最佳模型测试集loss性能变化，如图所示。
% 具体来说，通过历史25个时隙的CSI预测未来N个连续时隙的CSI，其预测时隙范围为1-25。我们将真实CSI和预测CSI的归一化均方误差(NMSE)作为测试集loss，分别对比ChannelGPT与所有基线在该预测任务下的步长与对应最佳模型测试集loss性能变化，如图所示。
Specifically, the historical CSI of 25 time slots is used to predict the future CSI of N consecutive time slots with a prediction slot size ranging from 1-20. To evaluate the model performance, the normalized mean square error (NMSE) of the true CSI and the predicted CSI are used as the evaluation metrics. We compare the NMSE performance of the ChannelGPT with baselines at different prediction steps in the test set, as shown in Fig. \ref{test_1_20_loss}. 
% 不同基线模型在该任务表现了不同的性能趋势，但在精度方面未能与ChannelGPT相媲美。在不同步长的预测任务中，ChannelGPT的NMSE均低于所有基线，尤其在中等步长（5到18）范围内表现优越，展示了大模型对信道中时序任务的强大建模和学习能力。
Different baseline models demonstrate varying performance trends on this task, but none match ChannelGPT in terms of accuracy. With different slot sizes, ChannelGPT consistently achieves lower NMSE than all baseline models, particularly excelling in the medium slot size (5 to 18) range. This highlights the powerful modeling capability of large model for CSI prediction. 

\subsection{Channel prediction based on multimodal information}
\subsubsection{Dataset construction}
\begin{figure*}[htbp]%
    \centering
    \subfloat[\textcolor{black}{}]{
        \label{original_RT_scenario}
        \includegraphics[width=0.5\linewidth]{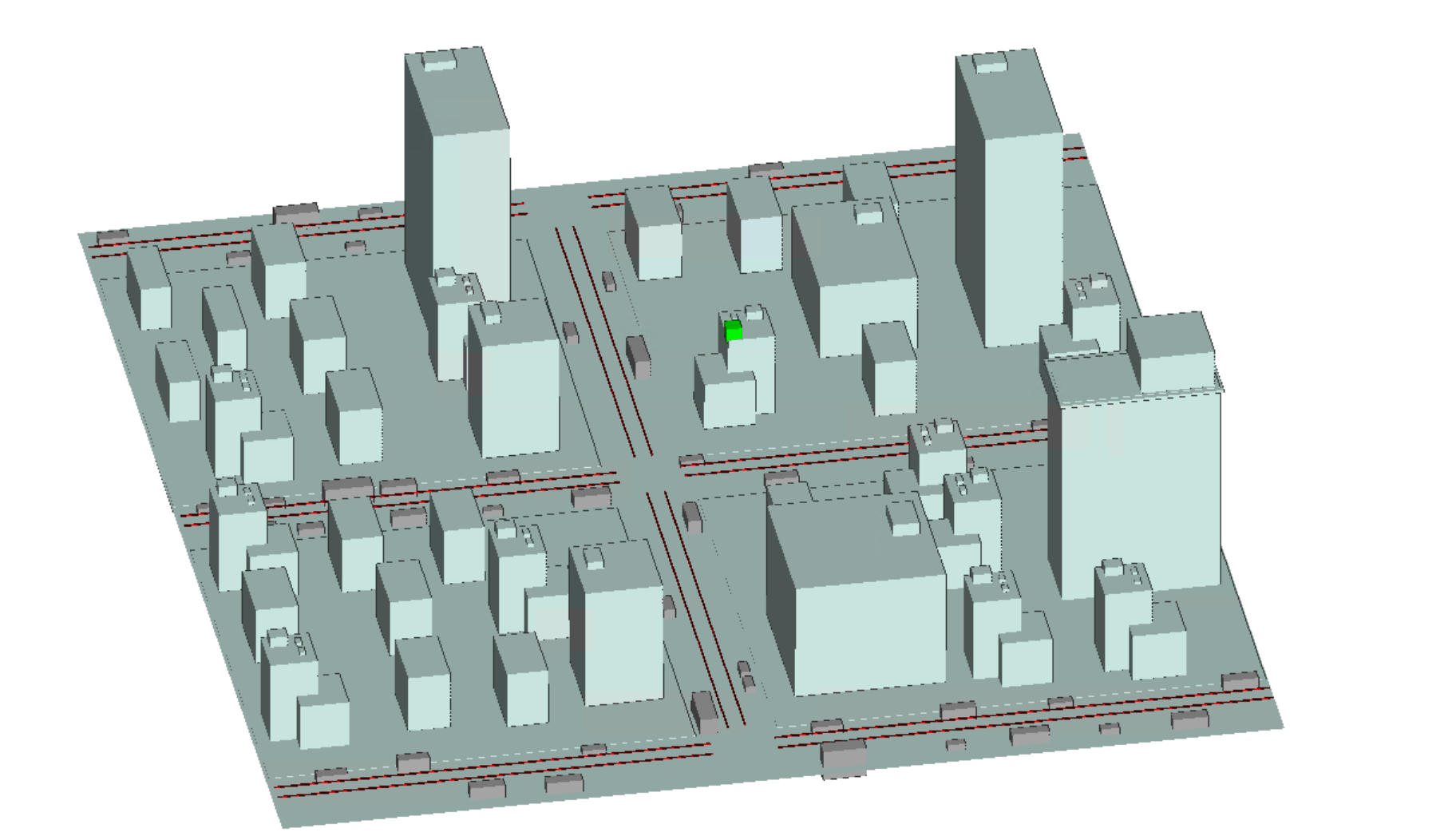}
        }
    \subfloat[\textcolor{black}{}]{
        \label{New_RT_scenario}
        \includegraphics[width=0.45\linewidth]{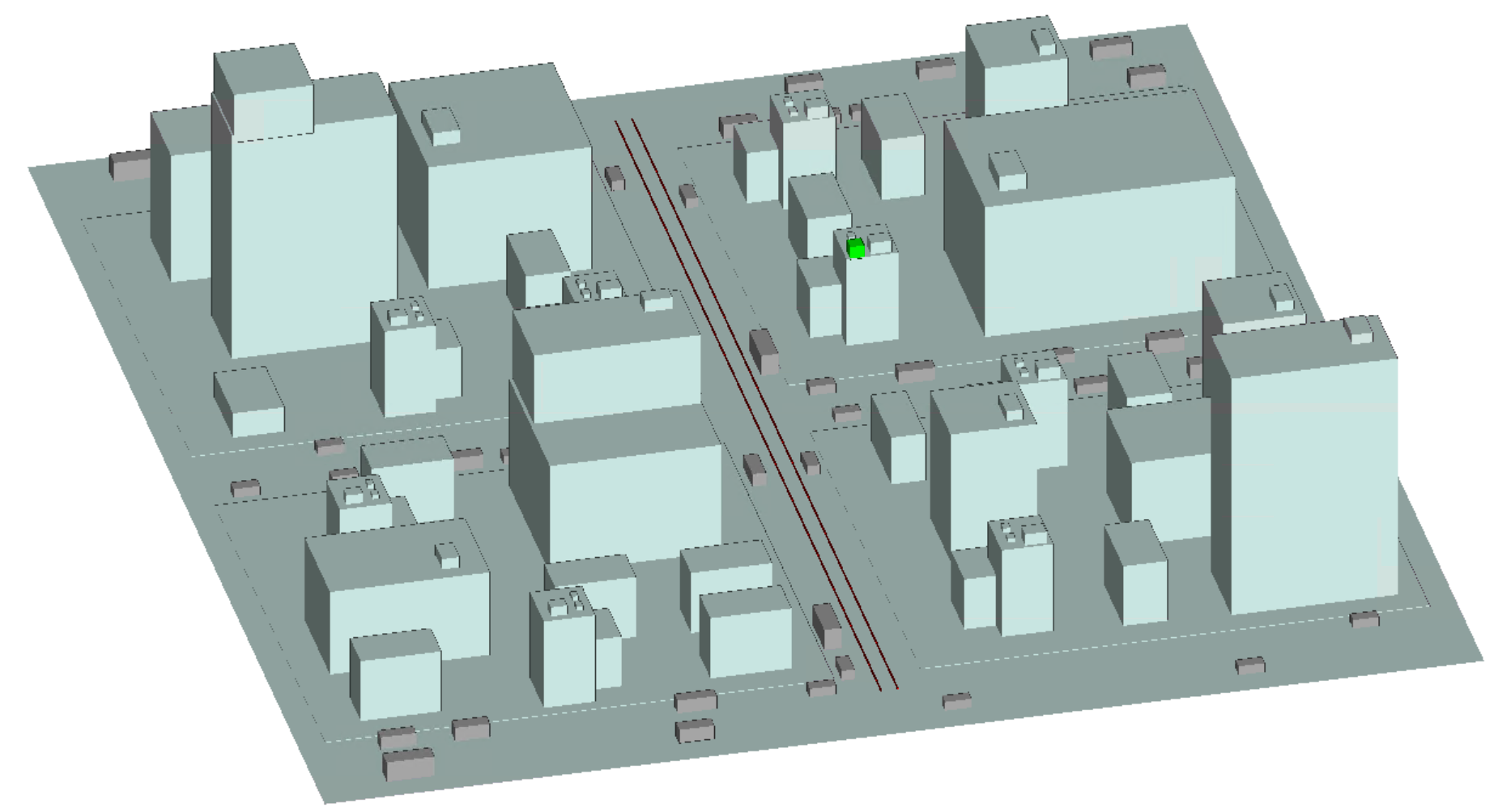}
        }\hfill
    \subfloat[\textcolor{black}{}]{
        \label{NMSE}
        \includegraphics[width=0.5\linewidth]{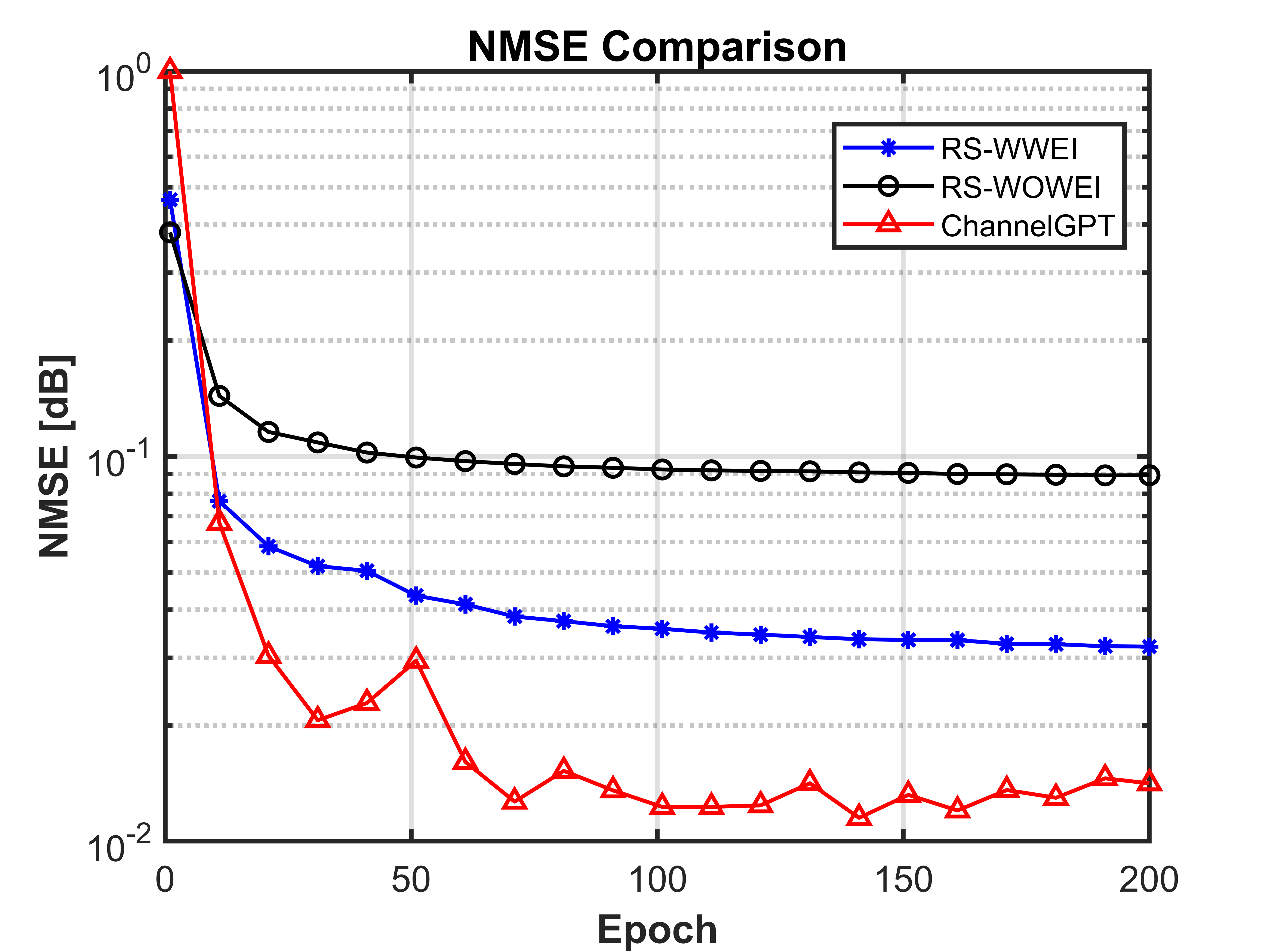}
        }
    \subfloat[\textcolor{black}{}]{
        \label{Cosine Similarity}
        \includegraphics[width=0.5\linewidth]{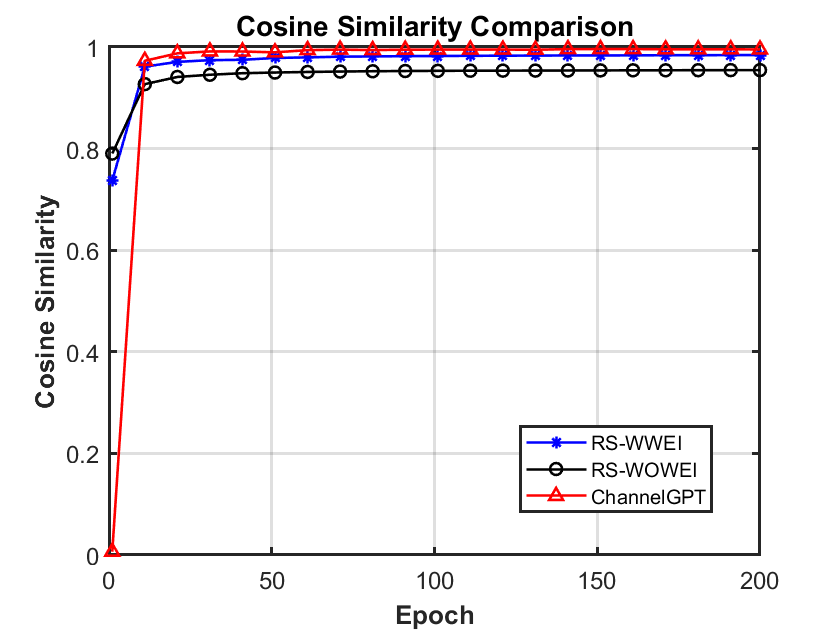}
        }
    \caption{The performance of ChannelGPT and other methods testing loss versus number of epochs, and the new scenario used for generalization validation. (a) The original scenario used to train models. (b) The new scenario. (c) NMSE. (d) Cosine Similarity.}
    \label{scenarios_performance_curve}
\end{figure*}

% 如图所示，我们构建了一个室外城市场景，长200m, 宽200m。该场景包含四栋建筑群和四条车道，以及不同类型的车辆随机放置在道路上，确保场景具有足够的多样性。该MISO系统包含配备UPA天线阵列的基站和配备单天线和多视角相机的移动用户，并采用OFDM调制，其中子载波个数为69，Tx放置在场景中间的建筑物顶部，一系列 Rx 放置在街道中间进行 RT 模拟，构建环境-信道多模态数据集。
As shown in Fig.\ref{scenarios_performance_curve} (a), we construct an outdoor urban scenario, 200m long and 200m wide. The scenario includes four building groups and four roads, with various types of vehicles randomly placed on the roads to ensure sufficient diversity. The multiple-input single-output (MISO) system consists of a BS equipped with a Uniform Planar Array (UPA) antenna and mobile users equipped with a single antenna and multi-view cameras. The system uses OFDM modulation with 69 subcarriers, where Tx is placed on top of a building in the center of the scene. A series of Rxs are placed along the streets to conduct RT simulations, constructing a multimodal environment-channel dataset. 
% \begin{figure*}[htbp]%
%     \centering
%     \subfloat[\textcolor{black}{}]{
%         \label{original_RT_scenario}
%         \includegraphics[width=0.5\linewidth]{PIC/old_scenario.png}
%         }
%     \subfloat[\textcolor{black}{}]{
%         \label{New_RT_scenario}
%         \includegraphics[width=0.45\linewidth]{PIC/new_scenario.png}
%         }\hfill
%     \subfloat[\textcolor{black}{}]{
%         \label{NMSE}
%         \includegraphics[width=0.5\linewidth]{PIC/NMSE.png}
%         }
%     \subfloat[\textcolor{black}{}]{
%         \label{Cosine Similarity}
%         \includegraphics[width=0.5\linewidth]{PIC/Corr.png}
%         }
%     \caption{The generalization performance of ChannelGPT with multimodal input. (a) The training scenario. (b) The testing scenario.  (c) The NMSE performance of the RS-WWEI and ChannelGPT testing loss versus number of epochs. (d) Cosine Similarity.}
%     \label{scenarios_performance_curve}
% \end{figure*}

\subsubsection{Model training and performance evaluation}
% 使用构建的环境-信道多模态数据集来训练模型，以减少导频资源开销。
The constructed environment-channel multimodal dataset is used to train the model, reducing pilot resource overhead. 
% 具体来说，random sampling with WEI (RS-WWEI)通过随机导频图案模式，并融合多视角图像数据和部分CSI数据来重建完整CSI。将网络设计分为三个模块，分别为：导频选择模块，多视角图像特征提取和基于特征融合的信道预测模块。random sampling with WEI and GPT2(ChannelGPT)前两个模块设计保持一致，第三个模块利用GPT2来重建完整CSI。random sampling without WEI (RS-WOWEI) 仅输入单模态部分CSI数据，去掉了多视角图像特征提取模块，其他保持一致。我们将RS-WWOEI作为基线。使用NMSE和余弦相似度来评估三种方法的性能。
Specifically, a random sampling with wireless environment information (RS-WWEI) fuses partial CSI data with multi-view image data to reconstruct the complete CSI. The network is divided into three modules: pilot selection module, multi-view image feature extraction module, and feature fusion-based channel prediction module. 
ChannelGPT remains the first two modules consistent, while the third module leverages ChannelGPT to reconstruct the full CSI. 
Random Sampling without WEI (RS-WOWEI) inputs only partial CSI data, omitting the multi-view image feature extraction module while keeping everything else consistent. 
RS-WOWEI serves as the baseline. 
NMSE and Cosine Similarity are used to assess the performance of the three methods.

% 精度：如图c,d所示，在导频比例均为1/8的情况下，ChannelGPT的初始收敛速度比其他方法更快。此外，训练后的ChannelGPT在NMSE和余弦相似度方面优于其他方法。
As shown in Fig.\ref{scenarios_performance_curve} (c) and Fig.\ref{scenarios_performance_curve} (d), with the number of 1/8 pilots, ChannelGPT demonstrates a faster initial convergence speed compared with other methods. Additionally, the trained ChannelGPT outperforms other methods in terms of NMSE and cosine similarity. 
% 泛化性：此外，为了探索大模型与信道任务结合的泛化能力，重新构建新场景，如图b所示，房屋和车辆的数量、位置与图a有很大差异。利用旧场景训练好的最优模型在新场景中按照两个指标来重新评估，如表所示。RS-WWEI和RS-WOWEI相较于旧场景的性能有所降低，而ChannelGPT在两个场景的性能表现基本一致，具有场景泛化能力。
In addition, to explore the generalization capability of large models in wireless channel tasks. A new scenario is constructed, as shown in Fig.\ref{scenarios_performance_curve} (b), where the number and positions of buildings and vehicles differ significantly from those in Fig.\ref{scenarios_performance_curve} (a). Trained on the original scenario, the optimal models are re-evaluated in the new scenario using NMSE and cosine similarity, as shown in Table \ref{new_scenario_comparison}. The performance of the RS-WWEI and RS-WOWEI is reduced compared to the original scenario, but ChannelGPT exhibits consistent performance in both scenarios, demonstrating strong generalization capability in diverse environments. 

\begin{table}[htbp]
\setlength{\tabcolsep}{11pt} %%%
\renewcommand\arraystretch{1.2}  %% 
\caption{\textcolor{black}{The model performance of comparison in new and origin scenarios.}}
\begin{center}
\fontsize{9}{9}\selectfont % 将字体大小设置为10pt，行间距设置为12pt
\begin{tabular}{|m{2cm}<{\centering}|m{0.7cm}<{\centering}|m{0.7cm}<{\centering}|m{0.7cm}<{\centering}|m{0.7cm}<{\centering}|}
\hline
 \multirow{2}{*}{Methods} & \multicolumn{2}{c|}{\textbf{Cosine Similarity}} & \multicolumn{2}{c|}{\textbf{NMSE}}\\
\cline{2-5}
 & {origin}  & {new} & {origin} & {new}\\
\hline
 {RS-WOWEI} & {0.9545}  & {0.9478} & {0.0893} & {0.1030}\\
\hline
 {RS-WWEI} & {0.9841}  & {0.9796} & {0.0320} & {0.0414}\\
\hline
 {ChannelGPT} & {0.9950}  & {0.9954} & {0.0128} & {0.0129}\\
\hline
\end{tabular}
\end{center}
\label{new_scenario_comparison}
\end{table}

\section{ChannelGPT challenges and future opportunities}
% 尽管前景广阔，ChannelGPT仍处于起步阶段，存在一些挑战，值得进一步研究。可分为三个主要挑战。 
Despite its promising future, ChannelGPT is still in its early stage and there are a number of challenges that warrant further research. They can be categorized into three main challenges. 

% 大规模环境与信道数据集构建：ChannelGPT依赖大规模、多模态的信道和环境数据进行建模。然而，如何有效、经济地采集、处理和存储这些数据，尤其是在动态的6G环境中，仍是一个巨大挑战。其次，分布式设备的数据同步也是亟需解决的问题，在不同节点间快速、精准地进行数据共享和同步，保证系统及时作出决策处理。此外，如何将通信领域的专有知识与大模型有效结合，如信道特性、电磁波传播机制等，仍需进一步研究。其中，REKP的深入研究是一个有潜力的课题。
\textbf{Large-scale environment and channel dataset collection: }ChannelGPT relies on extensive, multimodal datasets encompassing both channel and environment information for effective modeling. However, the challenge lies in how to efficiently and economically collect, process, and store massive multimodal data. Additionally, the synchronization of distributed devices is a challenging issue. Ensuring rapid and accurate data sharing and synchronization across different nodes is crucial to guarantee timely decision-making and system processing. Moreover, integrating specific knowledge from communications into large models remains an area that requires further exploration, such as channel characteristics, electromagnetic wave propagation mechanism, and so on. Notably, The REKP is a promising research direction.

% 与现有通信系统的接口集成融合：为了使ChannelGPT真正落地应用，其与现有通信系统的接口集成问题不容忽视。不同的通信协议、设备架构和标准化要求可能带来兼容性和互操作性的挑战。ChannelGPT可以与当前系统无缝连接，并在不引入操作差异的情况下增强其功能。
\textbf{Integration with existing communication system: }For ChannelGPT to be effectively deployed, integrating it with existing communication systems is a critical consideration. The diversity of communication protocols, device architectures, and standardization requirements may present challenges related to compatibility and interoperability. Addressing these challenges is essential to ensure that ChannelGPT can seamlessly interface with current systems and enhance their functionality without introducing operational discrepancies. 

% 硬件与能耗：大模型通常需要巨大的计算资源和能量消耗，这在6G网络中的大规模应用时会引发硬件负载和能效的矛盾。因此，如何在保持高性能的同时降低能耗，开发高效的硬件加速器和节能算法，成为ChannelGPT未来发展的重要方向。
\textbf{Hardware and energy consumption: }Large models typically require substantial computational resources and energy, especially when widely applied in 6G network, leading to a conflict between hardware load and energy efficiency. Therefore, the future development of ChannelGPT must address the challenge of reducing energy consumption while maintaining high performance. Developing efficient hardware accelerators and energy-saving algorithms will be key to enhancing the energy efficiency of ChannelGPT.

\section{Conclusion}
% 本文提出了ChannelGPT框架，是服务于DTC系统的智能核心。首先，分析了为什么6G需要ChannelGPT，与现有的一些方法进行了比较。阐述了DTC赋能6G网络的全流程，其中，6G和DTC交互必要的一环是环境智能。其次，详细阐述了ChannelGPT的框架，包括数据层，算法模型层，功能应用层，并解释了其五大核心能力。此外，我们还通过实验评估了大模型赋能通信任务的可行性，其结果展现了大模型的独特优势。最后，讨论了实现ChannelGPT的一些悬而未决的问题。
This article proposes the ChannelGPT framework as the intelligent core for the DTC system. Firstly, it analyzes the necessity of ChannelGPT for 6G and compares it with existing methods. The entire process of how DTC enables the 6G network is described, with EI acting as a crucial component of the interaction between 6G and DTC. Subsequently, the ChannelGPT framework is detailedly described, including the data processing layer, algorithm modeling layer, and functional application layer, with its five core capabilities. Additionally, we evaluate the feasibility of large models in enabling communication tasks through experiments, and the results demonstrate the advantages of accuracy, generalization and multimodal processing ability for ChannelGPT. Finally, some open issues and future opportunities related to the implementation of ChannelGPT are discussed.

\section{Acknowledgement}
This work is supported by the National Key R\&D Program of China (Grant No. 2023YFB2904805), the National Natural Science Foundation of China (No. 62401084), and BUPT-CMCC Joint Innovation Center. 

\bibliographystyle{IEEEtran}
\bibliography{cite}

\begin{IEEEbiographynophoto}{LI YU (li.yu@bupt.edu.cn)}
is currently a postdoctoral research fellow with Beijing University of Posts and Telecommunications.
\end{IEEEbiographynophoto}

\begin{IEEEbiographynophoto}{LIANZHENG SHI (shilianzheng@bupt.edu.cn)}
is currently pursuing the master’s degree with Beijing University of Posts and Telecommunications. 
\end{IEEEbiographynophoto}

\begin{IEEEbiographynophoto}{JIANHUA ZHANG (jhzhang@bupt.edu.cn) }
is currently a professor with Beijing University of Posts and Telecommunications.
\end{IEEEbiographynophoto}

\begin{IEEEbiographynophoto}{JIALIN WANG (wangjialinbupt@bupt.edu.cn)}
is currently pursuing the Ph.D. degree with Beijing University of Posts and Telecommunications. 
\end{IEEEbiographynophoto}

\begin{IEEEbiographynophoto}{ZHEN ZHANG (zhenzhang@imu.edu.cn)}
is currently a research fellow with Inner Mongolia University.
\end{IEEEbiographynophoto}

\begin{IEEEbiographynophoto}{YUXIANG ZHANG (zhangyx@bupt.edu.cn) }
is currently a associate researcher in Beijing University of Posts and Telecommunications.
\end{IEEEbiographynophoto}

\begin{IEEEbiographynophoto}{GUANGYI LIU (liuguangyi@chinamobile.com) }
is currently Fellow and 6G lead specialist, China Mobile Research Institute.
\end{IEEEbiographynophoto}
\end{document}